\documentstyle[preprint,aps,amsmath,tighten,graphicx]{revtex}

\begin{document}
\draft

\title{Low energy onset of nuclear shadowing in photoabsorption}

\author{T. Falter, S. Leupold and U. Mosel}

\address{Institut f\"ur Theoretische Physik\\ Justus-Liebig-Universit\"at
Giessen\\ D-35392 Giessen, Germany}

\date{February 26, 2000}

\maketitle

\begin{abstract}
The early onset of nuclear shadowing in photoabsorption at low
photon energies ($\gtrsim$ 1~GeV) has recently been interpreted as
a possible signature of a decrease of the $\rho$ meson mass in
nuclei. We show that one can understand this early onset within
simple Glauber theory if one takes the negative real part of the
$\rho N$ scattering amplitudes into account, corresponding to a
higher effective mass of the $\rho$ meson in nuclear medium.
\end{abstract}

\pacs{PACS numbers: 25.20.Dc, 24.10.Ht, 12.40.Vv}


Recent photoabsorption data \cite{Bia96,Muc98} on C, Al, Cu, Sn
and Pb in the energy range from 1 to 2.6~GeV display an early
onset of the shadowing effect. The shadowing of high energy
photons can be quantitatively understood in the Glauber approach
(see e.g. \cite{Bau78,Don78} and references therein) but some of
the newer models \cite{Pil95,Eng97} slightly underestimate the
effect for low photon energies or even predict antishadowing
\cite{Bof96} below 2~GeV when nucleon correlations are taken into
account.

The early onset of shadowing has recently \cite{Muc99} been
interpreted as a sign for a lighter $\rho$ meson in medium. The
shadowing effect was evaluated within a Glauber-Gribov multiple
scattering theory \cite{Gri70,Wei76} and generalized vector
dominance using realistic spectral functions for the hadronic
components of the photon but neglecting the real parts of the
hadron-nucleon scattering amplitudes. A decrease of the $\rho$
mass in nuclei was then suggested to fit the data.

The aim of this paper is to show how one can understand the data within
a simple Glauber model \cite{Gl59,Gl70,Yen71} if one takes the real
part of the $\rho N$ scattering amplitude into account. Experiments
\cite{Alv,Big} show that for energies of about 4 and 6~GeV the real
part of the $\rho N$ forward scattering amplitude is negative and
already of the same order of magnitude as its imaginary part.
Dispersion theoretical calculations \cite{Ele97,Kon98} indicate that
this is also the case for the energies we are considering. A negative
real part indeed leads to a positive mass shift of the $\rho$ in medium
as pointed out by Eletsky and Ioffe~\cite{Ele97}. We also include
two-body correlations between the nucleons which avoids unphysical
contributions to the shadowing effect in the considered energy region.

Glauber's formalism allows us to express the nuclear amplitudes of
high energy particles in terms of more fundamental interactions
with nucleons. We are interested in the total photon nucleus cross
section which is related to the nuclear forward Compton amplitude
via the optical theorem. We will  briefly describe the two
contributions to the nuclear Compton amplitude in order
$\alpha_{em}$. For a detailed derivation within the Glauber model we
refer to \cite{Yen71} and references therein.

\begin{figure}
\begin{center}
\includegraphics[width=5cm]{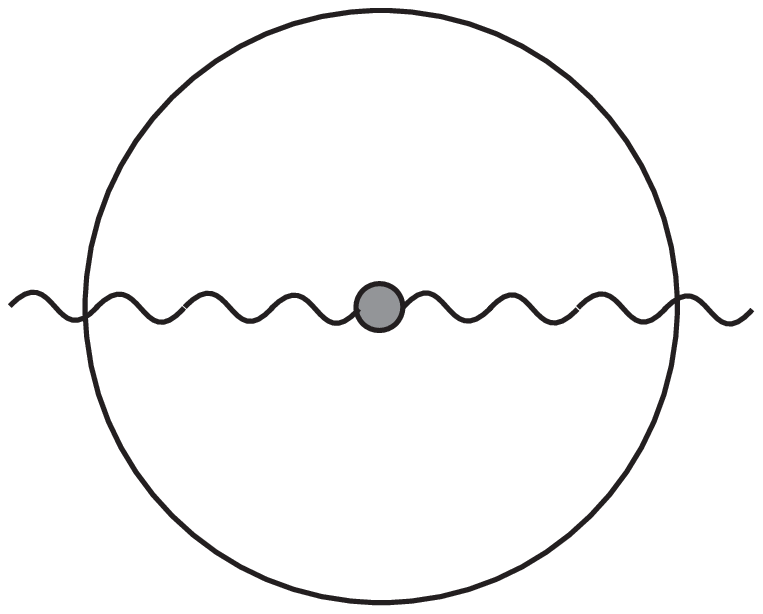}
\end{center}
\caption{Contributions to the unshadowed part of the nuclear
Compton amplitude in order $\alpha_{em}$. The photon scatters from
a single nucleon.} \label{fig:single}
\end{figure}

\begin{figure}
\begin{center}
\includegraphics[width=6.5cm]{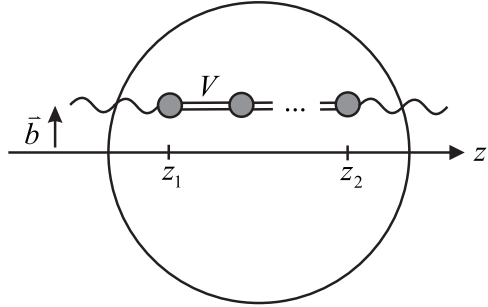}
\end{center}
\caption{Shadowing contribution to the nuclear Compton amplitude
in order $\alpha_{em}$. The photon produces a vector meson V that
scatters through the nucleus and finally back into the outgoing
photon.} \label{fig:multi}
\end{figure}

Consider a real photon with momentum $k\cdot\vec{e}_z$ that hits a
nucleus with mass number $A$ and nucleon number density
$n(\vec{r})$. The first contribution to the Compton amplitude in
order $\alpha_{em}$ comes from forward scattering of the photon
from a single nucleon somewhere inside the nucleus as shown in
Fig.~\ref{fig:single}. This leads to the unshadowed part of the
cross section (first term in (\ref{eq:cs})). The second
contribution comes from processes as depicted in
Fig.~\ref{fig:multi}. Here the photon enters the nucleus at impact
parameter $\vec{b}$ and produces an on-shell vector meson $V$ with
momentum $k_V\cdot\vec{e}_z$ on a nucleon at position $z_1$. This
meson then scatters at fixed impact parameter (eikonal
approximation) through the nucleus and finally at position $z_2$
back into the outgoing photon.

The resulting expression for the total photon nucleus cross
section, neglecting correlations between the single nucleons
(independent particle model), is then given by~\cite{Yen71}
\begin{equation}
\label{eq:cs}
  \begin{split}
    \sigma_{\gamma A}&=A\sigma_{\gamma N}+\underset{V}{\sum}\frac{8\pi^2}{kk_V}\textrm{Im}\biggl\{if_{\gamma V}f_{V\gamma}\int d^2b\int_{-\infty}^{\infty}dz_1\int_{z_1}^{\infty}dz_2n(\vec b,z_1)n(\vec b,z_2)\\
&\quad\times
e^{iq_V(z_1-z_2)}\exp\left[-\frac{1}{2}\sigma_V(1-i\alpha_V)\int_{z_1}^{z_2}dz'n(\vec
b,z')\right]\biggr\},
  \end{split}
\end{equation}
where $\sigma_{\gamma N}$ denotes the total photon nucleon cross section
and $\sigma_V$ the total vector meson nucleon cross section. The
momentum transfer $q_V=k-k_V$ is given by
\begin{equation}
\label{eq:qv} q_V=k-\sqrt{k^2-m_V^2}
\end{equation}
where $m_V$ is the vacuum mass of the vector meson $V$. We use the
vector dominance model (VDM) to relate the photoproduction
amplitude $f_{\gamma V}$ for the vector meson $V$ in the forward
direction to the $VN$ forward scattering amplitude $f_{VV}$ in the
following way
\begin{equation}
\label{eq:photoamp} f_{\gamma
V}f_{V\gamma}=\frac{e^2}{g_V^2}f_{VV}^2.
\end{equation}
The optical theorem allows us to express the amplitude $f_{VV}$ as
\begin{equation}
\label{eq:fvv}
f_{VV}=\frac{ik_V}{4\pi}\sigma_V(1-i\alpha_V)
\end{equation}
where $\alpha_V=\textrm{Re}f_{VV}/\textrm{Im}f_{VV}$ is the ratio of real to
imaginary part of the $VN$ forward scattering amplitude. From
(\ref{eq:photoamp}) one sees that within VDM the photoproduction
amplitude also depends on $\alpha_V$.

The effect of two-body correlations between the nucleons has been
investigated e.g. in Ref. \cite{Don78,Yen71}. We are interested in how
these correlations influence the shadowing in the low energy region
where $q_V$ is large. From (\ref{eq:cs}) we see that for very large
$q_V$ the second term on the right hand side contributes most if
$z_1\approx z_2$, that is when the first and the last nucleon in the
scattering process are approximately at the same position. Replacing
the product of one-particle densities by the two-particle density
\begin{equation}
\label{eq:corr}
  n_2(\vec b,z_1,z_2)=n(\vec b,z_1)n(\vec b,z_2)+\Delta(\vec{b},|z_1-z_2|)
\end{equation}
as proposed in Ref.~\cite{Bof96}, avoids such unphysical contributions.
Since for $z_1\approx z_2$ the last exponential in (\ref{eq:cs}) is
approximately one, consideration of correlations between the first and
the last nucleon should be sufficient. For the two-body correlation
function $\Delta$ we use the same Bessel function parametrization as in
Ref.~\cite{Bof96}:
\begin{equation}
\Delta(\vec{b},|z_1-z_2|)=-j_o(q_c|z_1-z_2|)n(\vec b,z_1)n(\vec
b,z_2)
\end{equation}
with $q_c$=780~MeV.

Since the largest contributions to shadowing stem from the lighter
vector mesons we only allow for $V=\rho,\omega,\phi$, neglecting higher
mass intermediate states with large $q_V$. Since the $\rho$ is the
lightest vector meson and its photoproduction amplitude
$f_{\gamma\rho}$ is about 3 times larger than that of the $\omega$ and
$\phi$, it will make the main contribution to the sum in (\ref{eq:cs})
for low energies. Hence, the shadowing effect at low photon energies is
very sensitive to the properties of the $\rho$ and in particular to the
choice of $\alpha_\rho$.

\begin{figure}
\begin{center}
\includegraphics[width=12cm]{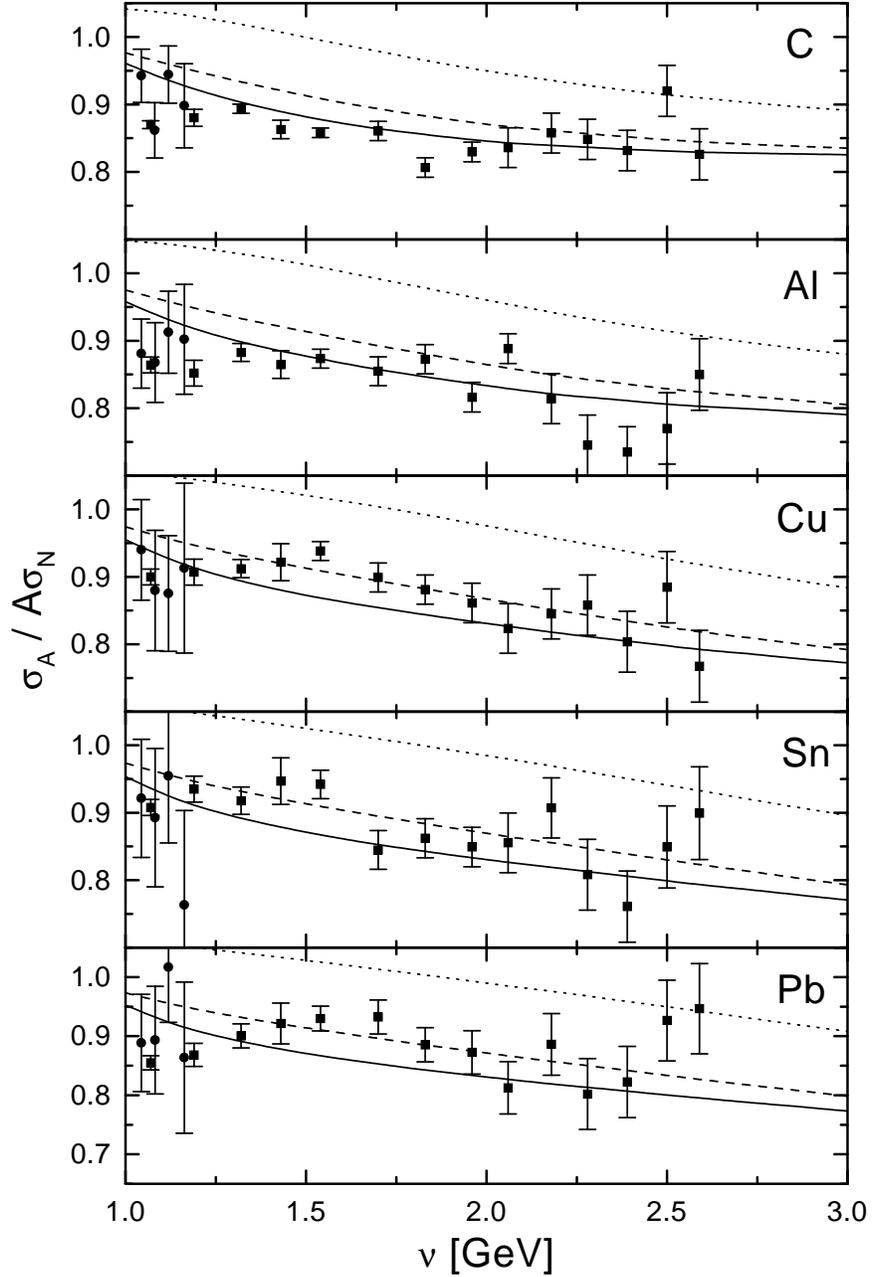}
\end{center}
\caption{Ratio of nuclear and nucleon photoabsorption cross
section plotted against the photon energy (dotted line -  real
part of the scattering amplitudes set to 0, dashed line - real
part like in Ref.~\protect\cite{Bau78}, solid line - $\rho N$
scattering amplitude like in Ref.~\protect\cite{Kon98}). The data
are taken from Ref.~\protect\cite{Bia96} (circles) and
Ref.~\protect\cite{Muc98} (squares).} \label{fig:aeff1}
\end{figure}

In Fig.~\ref{fig:aeff1} we compare the ratio $\sigma_{\gamma
A}/A\sigma_{\gamma N}$ plotted against photon energy with the data
from Ref.~\cite{Bia96,Muc98} for different nuclei. We assume a
Woods-Saxon distribution~\cite{Lenske} for $n(\vec{r})$ and
approximate the photon nucleon cross section $\sigma_{\gamma N}$
for each nucleus with mass number $A$ and proton number $Z$ by
\begin{equation}
\sigma_{\gamma N}=\frac{Z\sigma_{\gamma p}+(A-Z)\sigma_{\gamma
n}}{A},
\end{equation}
fitting the data on $\sigma_{\gamma p}$ and $\sigma_{\gamma n}$
for photon energies between 1 and 5~GeV.

The dotted line in Fig.~\ref{fig:aeff1} represents the result one
gets using the quark model parametrization for $\sigma_V$ and the
coupling constants $g_V$ of Model~I of Ref.~\cite{Bau78} with
$\alpha_V$ set to zero. One clearly underestimates the shadowing
effect at the considered energies and even gets antishadowing at
photon energies below 1.5~GeV ($^{12}\textrm{C}$) and 2~GeV ($^{208}\textrm{Pb}$) as stated in Ref.~\cite{Muc98,Bof96}.

The effect of the real part of $f_{\rho\rho}$ is shown by the dashed
line in Fig.~\ref{fig:aeff1}. The parametrization of $\alpha_V$ is also
taken from Model~I of Ref.~\cite{Bau78}\footnote{Due to a misprint the
sign of $\alpha_V$ in Table XXXV of Ref.~\cite{Bau78} is wrong.}.
Taking the negative real parts of the amplitudes into account leads to two
competing effects: The negative value of $\alpha_V$ in the exponential of (\ref{eq:cs})
diminishes the shadowing effect. However $\alpha_V$ also enters the prefactor
$f_{\gamma V}f_{V\gamma}$ via (\ref{eq:photoamp}) and (\ref{eq:fvv}) and
thereby increases shadowing.
In total this leads to an enhancement of the shadowing effect and 
improves the agreement with
experiment significantly. This result is also obtained in \cite{Ef99}
which investigates the influence of shadowing on photo-meson production
for photon energies between 1 and 10~GeV.

The solid line in Fig.~\ref{fig:aeff1} shows the result one gets if one
uses the $\rho N$ scattering amplitude from the dispersion theoretical
analysis by Kondratyuk et al~\cite{Kon98} and assuming that
$f_{\omega\omega}=f_{\rho\rho}$. For the $\phi$ we still use the
parametrization from Ref.~\cite{Bau78}. In Ref.~\cite{Kon98} the
dependence of $f_{\rho\rho}$ on the momentum of the $\rho$ meson yields
a positive mass shift for $\rho$ momenta larger than 100~MeV. This is
compatible with the result of Eletsky and Ioffe~\cite{Ele97} for
energies above 2~GeV who obtain an increase of the $\rho$ mass in
medium with growing $\rho$ momentum. Again one sees that considering
the (negative) real part of the $\rho N$ scattering amplitude leads to
a very good agreement with the data. The difference between the dashed
and the solid lines in Fig.~\ref{fig:aeff1} reflects the uncertainty in
the elementary $\rho N$ amplitude; it is, however, obvious that both
parametrizations used lead to the same conclusion that using a negative
real part for the $\rho N$ scattering amplitude explains the early
onset of shadowing.

\begin{figure}
\begin{center}
\includegraphics[width=6.5cm]{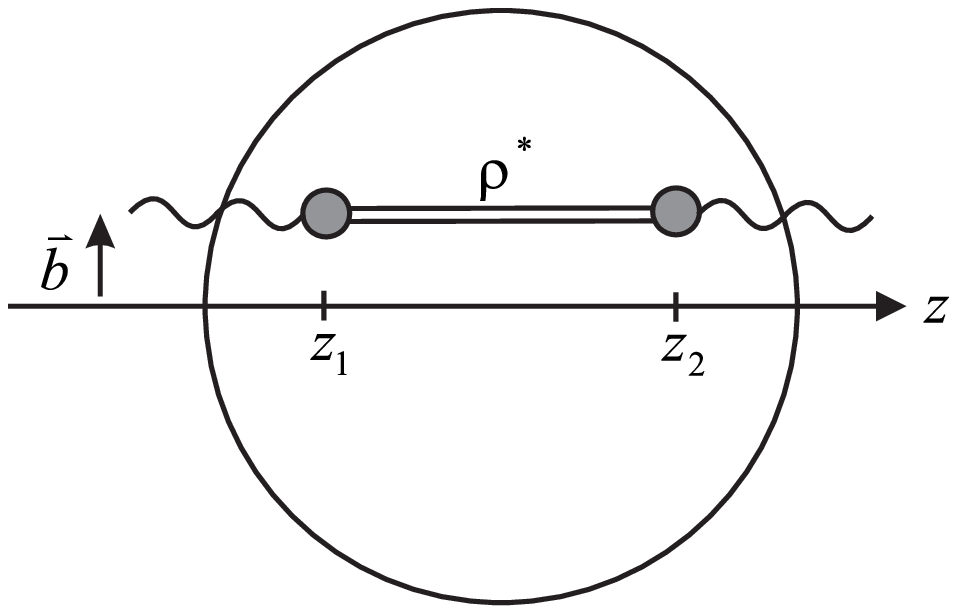}
\end{center}
\caption{The effective propagator of the $\rho^*$ replaces the multiple
scattering of the $\rho$ meson in Fig.~\ref{fig:multi}.}
\label{fig:effective}
\end{figure}

This result can be interpreted in terms of an in-medium change of the
$\rho$ meson properties. For large energies $k\approx k_\rho\gg m_\rho$
$q_\rho\approx m_\rho^2/2k_\rho$ the last two exponentials in
(\ref{eq:cs}) simplify to
\begin{equation*}
  \exp\left[\frac{i}{2k_\rho}\left(m_\rho^2-4\pi f_{\rho\rho}n_0\right)(z_1-z_2)\right]
\end{equation*}
where we have assumed a uniform density $n_0$. We would have gotten the
same result if we had taken for the $\rho$ contribution to the Compton
amplitude the process pictured in Fig.~\ref{fig:effective}. Here the
photon produces an effective $\rho^*$ with mass $m^*$ and width
$\Gamma^*$ at the first nucleon at position $z_1$ which propagates
formally without further scattering to the nucleon at position $z_2$
and scatters back into the photon. The effective propagator contains
the multiple scattering of the $\rho$ and can be calculated from the
effective optical potential
\begin{equation}
  U=-4\pi f_{\rho\rho}n_0.
\end{equation}
In the calculation reported here the negative real part of the
amplitude $f_{\rho\rho}$ results in a \emph{larger} effective mass of the
$\rho$ meson in medium
\begin{equation}
m^* = \sqrt{m_\rho^2 - 4\pi\textrm{Re}\left(f_{\rho\rho}\right) n_0}~.
\end{equation}
Thus, the multiple scattering contained in (\ref{eq:cs}) generates the
mass-shift of the $\rho$-meson. On the other hand, using an external
mass-shift for the $\rho$ meson in a multiple scattering approximation
such as (\ref{eq:cs}) doublecounts the in-medium effects. This explains
why the authors of \cite{Muc99} were led to the conclusion that the
early onset of shadowing reflects a \emph{lowering} of the $\rho$
meson mass in medium.

We have demonstrated that the early onset of shadowing can be
understood if one allows for a negative real part of the vector meson
scattering and photoproduction amplitudes.  This corresponds to an
increase of the $\rho$ meson mass in the nuclear medium at large
momenta, in agreement with dispersion theoretical analyses.

\acknowledgements
This work was supported by DFG.

\newpage


\begin{thebibliography}{99}

\bibitem{Bia96} N.~Bianchi {\it et al.}, Phys. Rev. C {\bf 54}, 1688 (1996).

\bibitem{Muc98} V.~Muccifora {\it et al.}, Phys. Rev. C {\bf 60}, 064616 (1999).

\bibitem{Bau78} T.~H.~Bauer, F.~Pipkin, R.~Spital and D.~R.~Yennie, Rev. Mod. Phys. {\bf 50}, 261 (1978).

\bibitem{Don78} A.~Donnachie and G.~Shaw, {\it Electromagnetic Interactions of Hadrons}, edited by A.~Donnachie and G.~Shaw (Plenum, New York, 1978), Vol. 2, p. 195.

\bibitem{Pil95} G.~Piller, W.~Ratzka, W.~Weise, Z. Phys. A {\bf 352}, 427 (1995).

\bibitem{Eng97} R.~Engel, J.~Ranft and S.~Roesler, Phys. Rev. D {\bf 55}, 6957 (1997).

\bibitem{Bof96} S.~Boffi, Ye.~Golubeva, L.~A.~Kondratyuk and M.~I.~Krivoruchenko, Nucl. Phys. A {\bf 606}, 421 (1996).

\bibitem{Muc99} N.~Bianchi, E.~De Sanctis, M.~Mirazita and V.~Muccifora, Phys. Rev. C {\bf 60}, 064617 (1999).

\bibitem{Gri70} V.~N.~Gribov, JETP {\bf 30}, 709 (1970).

\bibitem{Wei76} J.~H.~Weis, Acta Phys. Pol. B {\bf 7}, 851 (1976).

\bibitem{Gl59} R.~J.~Glauber, {\it Lectures in Theoretical Physics}, edited by W.E.~Brittin and L.G.~Dunham (Wiley Intersience, New York, 1959), Vol. I, p. 315.

\bibitem{Gl70} R.~J.~Glauber, {\it High Energy Physics and Nuclear Structure}, edited by S.~Devons (Plenum, New York, 1970), p. 207.

\bibitem{Yen71} D.~R.~Yennie, {\it Hadronic Interactions of Electrons and Photons}, edited by J.~Cummings and H.~Osborn (Academic, New York/London, 1971), p. 321.

\bibitem{Alv} H.~Alvensleben {\it et al.}, Phys. Rev. Lett. {\bf 25}, 1377 (1970); Phys. Rev. Lett. {\bf 27}, 444 (1971).

\bibitem{Big} P.~J.~Biggs, D.~W.~Braben, R.~W.~Clifft, E.~Gabathuler and R.~E.~Rand, Phys. Rev. Lett. {\bf 27}, 1157 (1971).

\bibitem{Ele97} V.~L.~Eletsky and B.~L.~Ioffe, Phys. Rev. Lett.
{\bf 78}, 1010 (1997).

\bibitem{Kon98} L.~A.~Kondratyuk, A.~Sibirtsev, W.~Cassing, Ye.~S.~Golubeva and M.~Effenberger, Phys. Rev. C {\bf 58}, 1078 (1998).

\bibitem{Lenske} H.~Lenske (private communication).

\bibitem{Ef99} M.~Effenberger and U.~Mosel, nucl-th/9908078.


\end{thebibliography}
\end{document}